\documentclass[aps,amsmath,amssymb,physrev,reprint,superscriptaddress]{revtex4-2}

\usepackage{siunitx}
\usepackage{graphicx}
\usepackage[normalem]{ulem}
\usepackage{xcolor}
\usepackage{multirow}

\newcommand\nlsout{\bgroup\markoverwith{\textcolor{green}{\rule[0.5ex]{2pt}{0.4pt}}}\ULon}

\begin{document}


\title{Dielectric perturbations: anomalous resonance frequency shifts in optical resonators}


\author{Farhan Azeem}
\author{Luke S. Trainor} 
\author{Patrick A. Devane}
\author{Daniel S. Norman}
\author{Alfredo Rueda}
\author{Nicholas J. Lambert}
\author{Madhuri Kumari}
\affiliation{The Dodd-Walls Centre for Photonic and Quantum Technologies, New Zealand}
\affiliation{Department of Physics, University of Otago, 730 Cumberland Street, Dunedin 9016, New Zealand}
\author{Matthew R. Foreman}
\affiliation{Blackett Laboratory, Department of Physics, Imperial College London, Prince Consort Road, London SW7 2AZ, UK}
\author{Harald G. L. Schwefel}
\email{harald.schwefel@otago.ac.nz}
\affiliation{The Dodd-Walls Centre for Photonic and Quantum Technologies, New Zealand}
\affiliation{Department of Physics, University of Otago, 730 Cumberland Street, Dunedin 9016, New Zealand}

\date{\today}

\begin{abstract}
Small perturbations in the dielectric environment around a high quality whispering gallery mode resonator usually lead to a frequency shift of the resonator modes directly proportional to the polarizability of the perturbation. Here, we report experimental observations of strong frequency shifts that can be opposite and even exceed the contribution of the perturbations' polarizability. The mode frequencies of a lithium niobate whispering gallery mode resonator are shifted using substrates of refractive indices ranging from 1.50 to 4.22. Both blue- and red-shifts are observed, as well as an increase in mode linewidth, when substrates are moved into the evanescent field of the whispering gallery mode. We compare the experimental results to a theoretical model by Foreman et al.~\cite{foreman_dielectric_2016} and provide an additional intuitive explanation based on the Goos-Hänchen shift for the optical domain. 
\end{abstract}


\maketitle

\section{Introduction}   
 
Small perturbations are at the core of many branches of physics. The textbook description of a small perturbation in the permittivity or permeability within an electromagnetic resonant system, leads to the Bethe-Schwinger perturbation theory~\cite{waldron_perturbation_1960}. One of its most straightforward applications is to the case of a dielectric microresonator perturbed by a contribution to the dielectric environment for which it predicts a small shift of the resonance frequency. If the perturbation is small enough it was assumed that the dipole approximation is valid, whereby the complex frequency shift is related to the polarizability and hence results in a red-shift of the resonance frequency. This view was particularly fruitful in considering optical microresonators for sensing of refractive index~\cite{teraoka_perturbation_2003,xiao_coupled_2008}, nanoparticles~\cite{arnold_whispering_2009,baaske_optical_2012} such as viruses~\cite{vollmer_single_2008}, bacteria~\cite{ren_high-q_2007}, single molecules~\cite{vollmer_single_2008} and proteins~\cite{vollmer_protein_2002, santiago-cordoba_nanoparticle-based_2011}, as well as in thermal sensing~\cite{dong_fabrication_2009}. Recently, however more in-depth studies have shown that small perturbations can have much more complex results, i.e., both red- and blue-shifts of the resonance frequency are possible~\cite{ruesink_perturbing_2015,foreman_dielectric_2016}. The key factor herein is that in an open system the radiative contributions of the perturbation---besides the mode broadening~\cite{sedlmeir_polarization-selective_2017}---need to be taken into account as well. This has lead to the experimental discovery of strong blue-shifts in the perturbation of a small toroidal resonator with plasmonic antennas~\cite{ruesink_perturbing_2015}.

Another prediction of this theory is that such shifts also exist when a dielectric interface is introduced into the vicinity of a dielectric whispering gallery mode (WGM) resonator~\cite{foreman_dielectric_2016}. Optical WGM resonators are based on the principle of total internal reflection (TIR) of light along the internal boundary of a transparent dielectric~\cite{strekalov_nonlinear_2016}. In low loss materials quality ($Q$) factors of over one hundred million have readily been achieved in varying geometries~\cite{sedlmeir_experimental_2013,lin_high-q_2012}, leading to applications in optical sensing~\cite{sedlmeir_high-q_2014,foreman_whispering_2015}, and when an optically non-linear material is used~\cite{breunig_three-wave_2016}, resulting in generation of frequency combs~\cite{kippenberg_microresonator-based_2011, rueda_resonant_2019}. An extended list of applications of WGM resonators also includes non-linear photonics~\cite{lin_nonlinear_2017,trainor_selective_2018}, quantum optics~\cite{strekalov_nonlinear_2016,otterpohl_squeezed_2019,fortsch_near-infrared_2015,shafiee_nonlinear_2020}, microwave to optical conversion~\cite{rueda_efficient_2016,lambert_coherent_2020,hease_bidirectional_2020}, optical communication~\cite{eschmann_stability_1994,monifi_tunable_2013,pfeifle_coherent_2014}, electro-optical modulation~\cite{yuce_optical_2009}, and lasers~\cite{he_whispering_2013}.

The effect of dielectric perturbations has already been used to carefully phase-match parametric down conversion in lithium niobate WGM resonators~\cite{schunk_interfacing_2015}. Various other methods are also employed to shift the resonance position in WGM resonators, including exploiting the thermo-optical~\cite{ward_thermo-optical_2010,vogt_thermal_2018} and electro-optical effects~\cite{savchenkov_tunable_2003}, as well as deforming the resonator by external strain~\cite{ilchenko_strain-tunable_1998,ioppolo_pressure_2007}.

\begin{figure*}
\centering
\includegraphics[width=\linewidth]{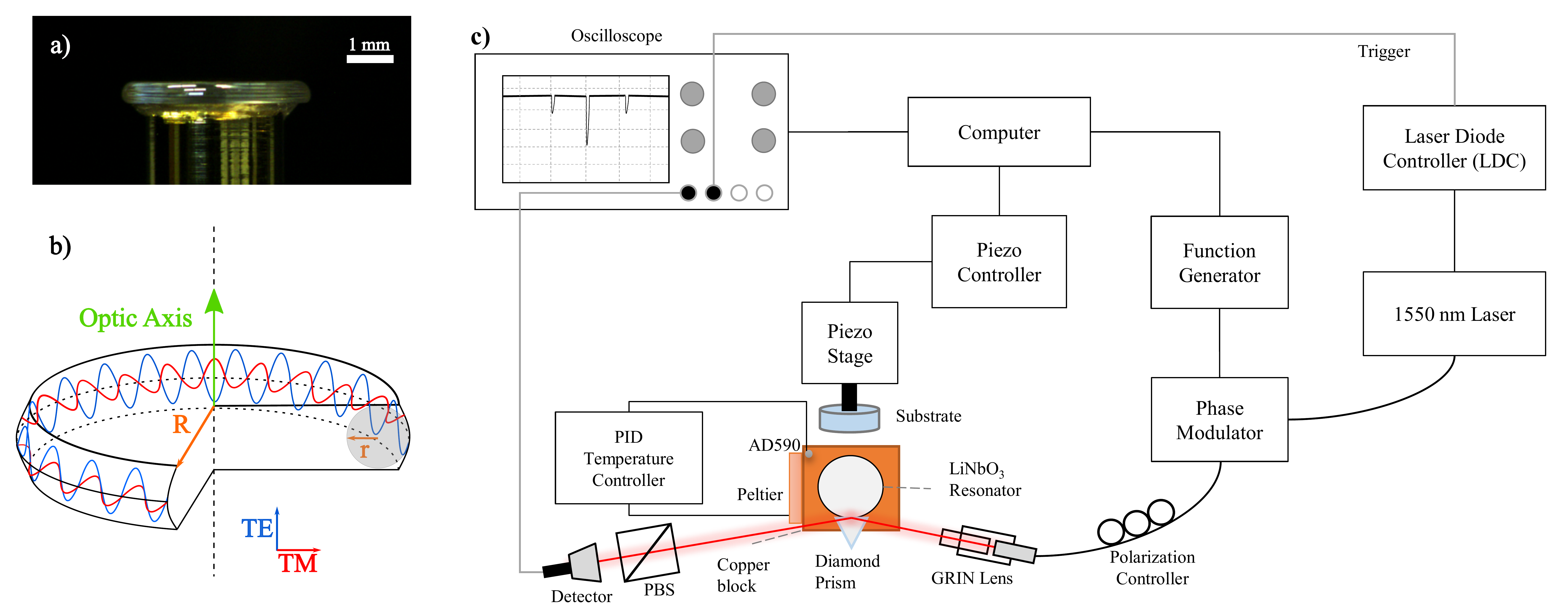}
\caption{Configuration of the experiment. a) The  $z$-cut LiNbO$_3$ WGM resonator is shown on the top left.  b) Schematic of the WGM showing the optic axis, the polarization of the modes, the radius of the disk $(R)$ and the radius of curvature of the resonator $(r)$. c) Schematic of the experimental setup to observe the effect of planar dielectric substrates on the WGMs of LiNbO$_3$ resonator. A telecom laser is evanescently coupled by using a diamond prism into a temperature stabilized resonator. The input polarization is controlled and sidebands can be modulated onto the laser for calibration. Both the prism and the dielectric substrate are moved using piezo positioners with nanometer precision.}\label{fig:Schematic}
\end{figure*}

Frequency tuning of WGMs by perturbing the dielectric environment has been studied analytically~\cite{foreman_dielectric_2016} for a disk-shaped lithium niobate (LiNbO$_3$) resonator. The results of this study show mode broadening in both the transverse electric (TE) and transverse magnetic (TM) polarized modes. The TM polarized modes exhibit only a red-shift in frequency for the given range of refractive indices studied, whereas the TE polarized modes exhibit both a red-shift and an anomalous blue-shift. 
Moreover, an anomalous blue-shift in the mode frequency of a toroid-shaped silica resonator has been demonstrated experimentally~\cite{ruesink_perturbing_2015}, via gold nano-antennas deposited on a glass substrate. Furthermore, blue-shifts have been experimentally demonstrated in a silicon sphere WGM resonator in the terahertz (THz) domain as well~\cite{vogt_anomalous_2019}. 

Here, for the first time we show experimental results confirming the prediction in Ref.~\cite{foreman_dielectric_2016}, in the optical domain. The frequency shifts and broadening of the mode are observed for both the TE and TM polarization. An anomalous blue-shift is observed for the TE polarization, when substrates of higher refractive index are brought into close proximity to the resonator. We also compare the results to an intuitive explanation based on the Goos-Hänchen shift.

\section{Experimental Setup}

 At the core of the experiment is a disk-shaped $z$-cut LiNbO$_3$ WGM resonator ($n_o=2.21$, $n_e=2.14$~\cite{zelmon_infrared_1997}), shown in Fig.~\ref{fig:Schematic} a). The symmetry of the resonator is shown in Fig.~\ref{fig:Schematic} b). The resonator has a major radius $R=\SI{2}{\mm}$ and a minor radius $r=\SI{0.23}{\mm}$, and modes with either TE or TM polarization can be excited. The schematic of the experiment is shown in Fig.~\ref{fig:Schematic} c). A fibre-coupled tunable diode laser (DL) with a central wavelength of \SI{1550}{\nano\meter} is the excitation source for the WGMs. The output from the DL is fed to a phase modulator (PM) to modulate sidebands onto the input laser signal for accurate frequency calibration of the DL scan range. A frequency scan of the laser reveals the reflection spectrum of the WGM. TE and TM modes are selected using a polarization controller. A diamond prism ($n_\text{p}=2.38$~\cite{phillip_kramers-kronig_1964}) is used to couple into the temperature stabilized resonator, with the laser output from the fibre focused onto the prism-resonator interface using a graded-index (GRIN) lens. The output light from the resonator is passed through a polarizing beam splitter (PBS) to select the desired polarization. The light is then collected by an indium gallium arsenide photodiode, and the output digitized using an oscilloscope.

A piezo stage moves the substrates into close proximity to the WGM resonator with nanometre accuracy. Seven substrates (listed in Table~\ref{tab:Substrates}) of varying refractive indices ($n_\text{sub}$)~\cite{amotchkina_characterization_2020,li_refractive_1980,marple_refractive_1964,debenham_refractive_1984,zelmon_infrared_1997,malitson_refraction_1962,malitson_refractive-index_1972,kim_absolute_2010,ghosh_temperature-dependent_1994} are used. The substrate is slowly moved towards the resonator until they are in contact, with a spectrum recorded at each step. Lorentzian fits are applied to the acquired spectra to extract the linewidths and the frequency positions of the mode. We took particular care that the same high $Q$ mode was used for all substrates.

\begin{table}[h]
\caption {Substrates with varying refractive index ($n_\text{sub}$) used to observe the frequency shifts of WGM resonances.}\label{tab:Substrates}
\begin{center}
\begin{tabular} {  l l l c } 
Substrate &Symbol &{$n_\text{sub}$}& Reference \\ 
\hline\hline
Germanium &Ge &4.22 &\cite{amotchkina_characterization_2020}\\ 
Silicon &Si &3.48 &\cite{li_refractive_1980}\\ 
Zinc Selenide &ZnSe &2.45 &\cite{marple_refractive_1964,amotchkina_characterization_2020}\\
Zinc Sulphide &ZnS &2.27 &\cite{debenham_refractive_1984,amotchkina_characterization_2020}\\ 
Lithium Niobate& \multirow{2}{*}{LN(o)}& \multirow{2}{*}{2.21} & \multirow{2}{*}{\cite{zelmon_infrared_1997}}\\ 
 (ordinary) & & &\\
Lithium Niobate&  \multirow{2}{*}{LN(e)}& \multirow{2}{*}{2.14} &  \multirow{2}{*}{\cite{zelmon_infrared_1997}}\\
(extra-ordinary) & & &\\ 
Sapphire& \multirow{2}{*}{Al$_2$O$_3$(o)} & \multirow{2}{*}{1.75} &  \multirow{2}{*}{\cite{malitson_refraction_1962,malitson_refractive-index_1972}}\\
 (ordinary) & & &\\ 
Sapphire& \multirow{2}{*}{Al$_2$O$_3$(e)} &\multirow{2}{*}{1.74} &\multirow{2}{*}{\cite{malitson_refractive-index_1972}}\\
 (extra-ordinary) & & &\\ 
BK7 Optical Glass &BK7 &1.50 &\cite{kim_absolute_2010,ghosh_temperature-dependent_1994}\\
\hline\hline
\end{tabular}
\end{center}
\end{table}

\section{Results and Discussion}
\begin{figure}[t!]
\centering
\includegraphics[width=\linewidth]{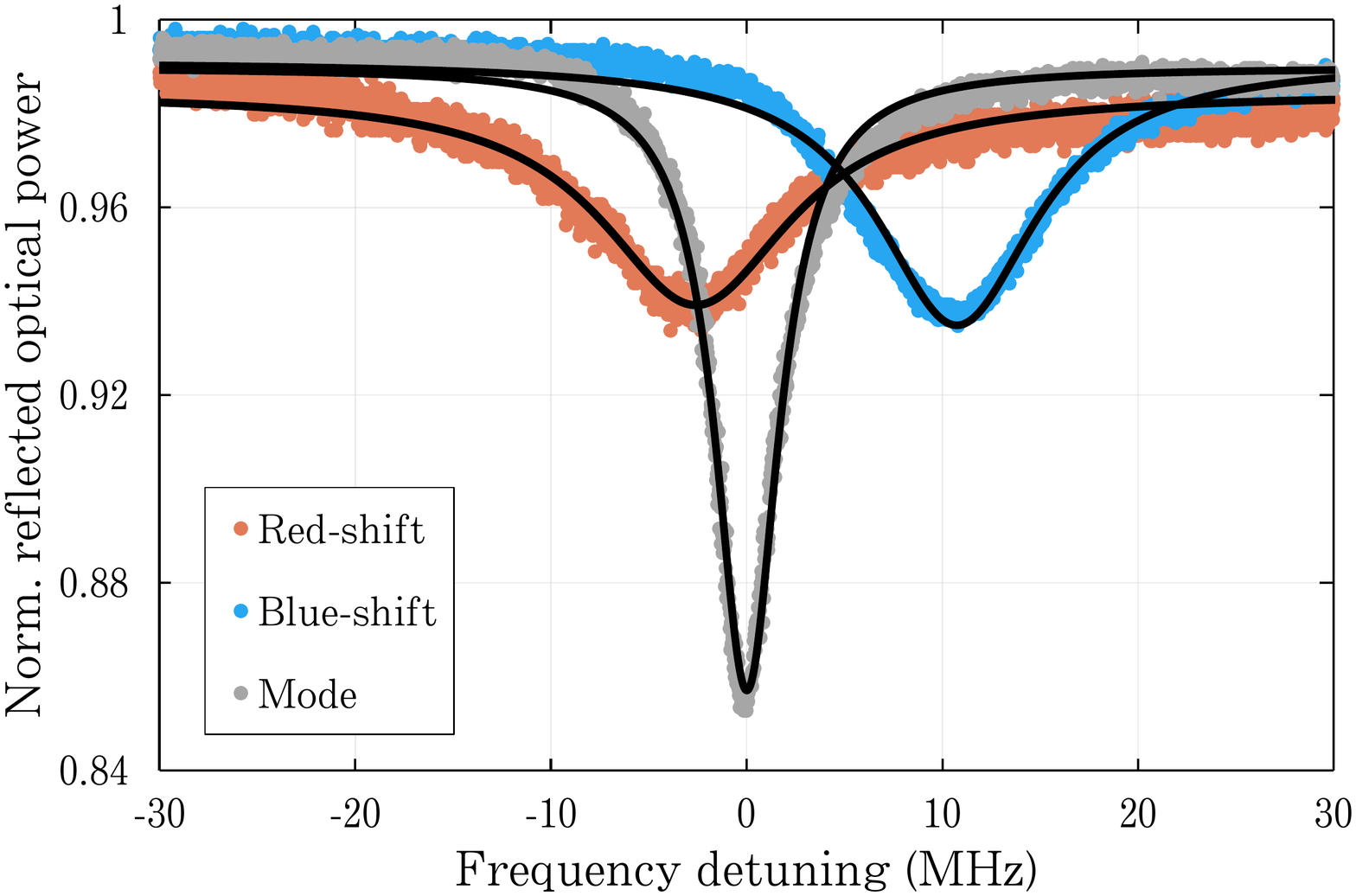}
\caption{Normalized reflected optical power versus frequency. The high $Q$ TE mode is shown in grey, whereas the red-shifted mode is depicted in red- and the blue-shifted mode in blue. The red-shift shown in the figure occurs due to the presence of LiNbO$_3$ substrate and the blue-shift due to Ge substrate within the evanescent field of the resonator. Mode broadening can also be observed in both instances. The coupling efficiency is reduced by more than 50\% in both the cases, which is mainly attributed to the additional out-coupling losses induced by the presence of the substrate.}\label{fig:RedBlueShift}

\end{figure}

As soon as the substrate penetrates a significant portion of the evanescent field of the mode a shift in the resonance frequency and mode broadening can be observed. 
Example red- and blue-shifts of the TE mode are plotted in Fig.~\ref{fig:RedBlueShift}. 
If the refractive index of the substrate is higher than that of the resonator $(n_\text{sub}>n_\text{res})$ the linewidth of the mode increases as the substrate approaches the resonator. 
Moreover, a decrease or an increase in the resonance frequency is observed depending on the type of substrate and polarization. 
The coupling efficiency to the mode drops as the distance between the resonator and the substrate decreases; this is mainly due to the additional losses induced by bringing the dielectric substrates within the evanescent field of the resonator. 

\begin{figure*}[t!]
\centering
\includegraphics[width=\linewidth]{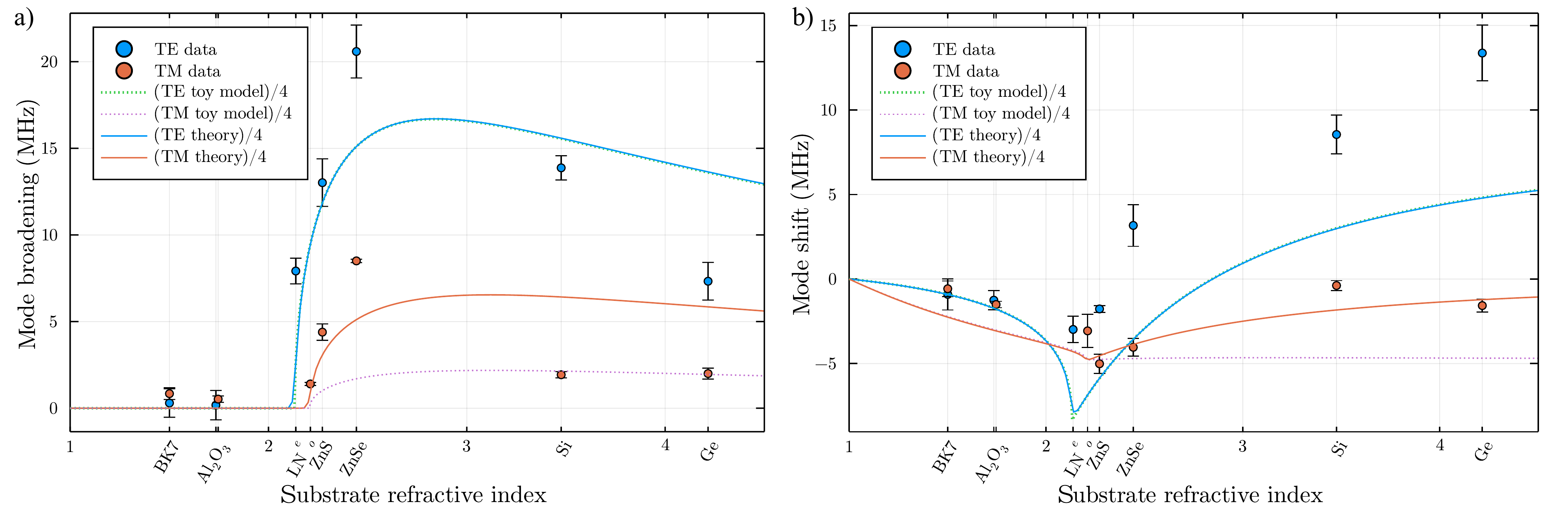}
\caption{The broadening of linewidth a) and resonance frequency shift b) in both TE and TM modes versus the refractive index of the substrates. The error-bars correspond to the 95 percent confidence interval deduced from repeated measurements. The solid curves depict the analytical theory of Foreman et al.~\cite{foreman_dielectric_2016}, the dotted curves the toy model of section~\ref{sec:toy}, each of which have been scaled by a factor of $1/4$.}\label{fig:ShiftBroadening}
\end{figure*}

To quantify this broadening we repeated the approach of the dielectric substrate to the resonator multiple times. A constant linear drift of the resonance frequency was removed as it was attributed to slow thermal heating/cooling of the device. After the removal, the exponential broadening of the linewidth and shift in the resonance frequency was recorded. The saturation point indicated the contact with the resonator and the final results from different runs was averaged. The increase in the linewidth of the mode due to each dielectric substrate is shown in Fig.~\ref{fig:ShiftBroadening} a).

Figure~\ref{fig:ShiftBroadening} b) shows the shift in the resonant frequency of the mode against the refractive index of the substrate. The TM polarized mode only exhibits the red-shift. However, the TE polarized mode shows both the red-shift and blue-shift. Dielectric substrates with a refractive index significantly larger than that of the resonator excite an anomalous blue-shift.  A blue-shift as high as \SI{14}{\mega\Hz} is observed in TE polarization, whereas a red-shift as low as \SI{-5}{\mega\Hz} is observed in TM polarization. The lowest values of the red-shift result with the ZnS substrate in the TM polarized mode, and with the LiNbO$_3$ substrates in the TE polarized mode. For LiNbO$_3$ the $n_\text{res}$ for TE polarization is the extraordinary refractive index (LN(e)), whereas for TM polarization it is the ordinary refractive index (LN(o)). 

Our results are in qualitative agreement with the theory described by Foreman et al.~\cite{foreman_dielectric_2016}. However, the model predicts both frequency shifts and mode broadening to be approximately a factor of four larger than those observed. We ascribe this discrepancy to experimental deviations from the ideal situation. In particular, the relevant surfaces of the substrates used to tune the resonances are unlikely to be completely parallel to the symmetry axis of the resonator. This leads to a small air gap between the resonator and substrate at the equator of the WGM, which reduces the magnitude of the effects due to the substrate. A small angle of \SI{2}{\degree} would lead to an air gap of \SI{140}{\nano\m}; this is comparable to the evanescent decay length of approximately  \SI{130}{\nano\m}.

\section{Toy model}\label{sec:toy}
The shape of the expected resonance shifts can also be explained by an altered Goos-Hänchen shift at the resonator rim, which causes the effective perimeter of the resonator to shift in or out.
When light undergoes TIR, it appears to shift laterally, a phenomenon termed the Goos-Hänchen shift~\cite{schwefel_direct_2008,bliokh_gooshanchen_2013} which has been successfully used in the analysis of deformed microresonators~\cite{unterhinninghofen_goos-hanchen_2008}. The reason is that the Fresnel reflection coefficient for TIR has an extra phase shift, $r_0 = -\exp(-i\Theta)$, such that the light appears to reflect off a surface that is a distance of $\delta R = \Theta/(2k_0 n \cos\theta_i)$ away from the real surface \cite{demchenko_analytical_2013,gorodetsky_geometrical_2006}, where $k_0$ is the vacuum wavevector of the light, $n$ is the refractive index, and $\theta_i$ is the angle of incidence of the light on the boundary. The angle of incidence can be approximated by $\cos\theta_i\approx\sqrt{-\zeta_q}(m/2)^{-1/3}$, where $m$ is the azimuthal mode number, $q$ is the radial mode number, and $\zeta_q$ is the $q$th root of the Airy function: $\mathrm{Ai}(\zeta_q) = 0$.

When we introduce a dielectric close to the resonator, we alter the phase of the reflection coefficient (and if the light can tunnel out into that dielectric, we alter its magnitude). We therefore alter the effective radius of the resonator in the region of the dielectric.

\begin{figure}[t!]
\centering
\includegraphics[width=\linewidth]{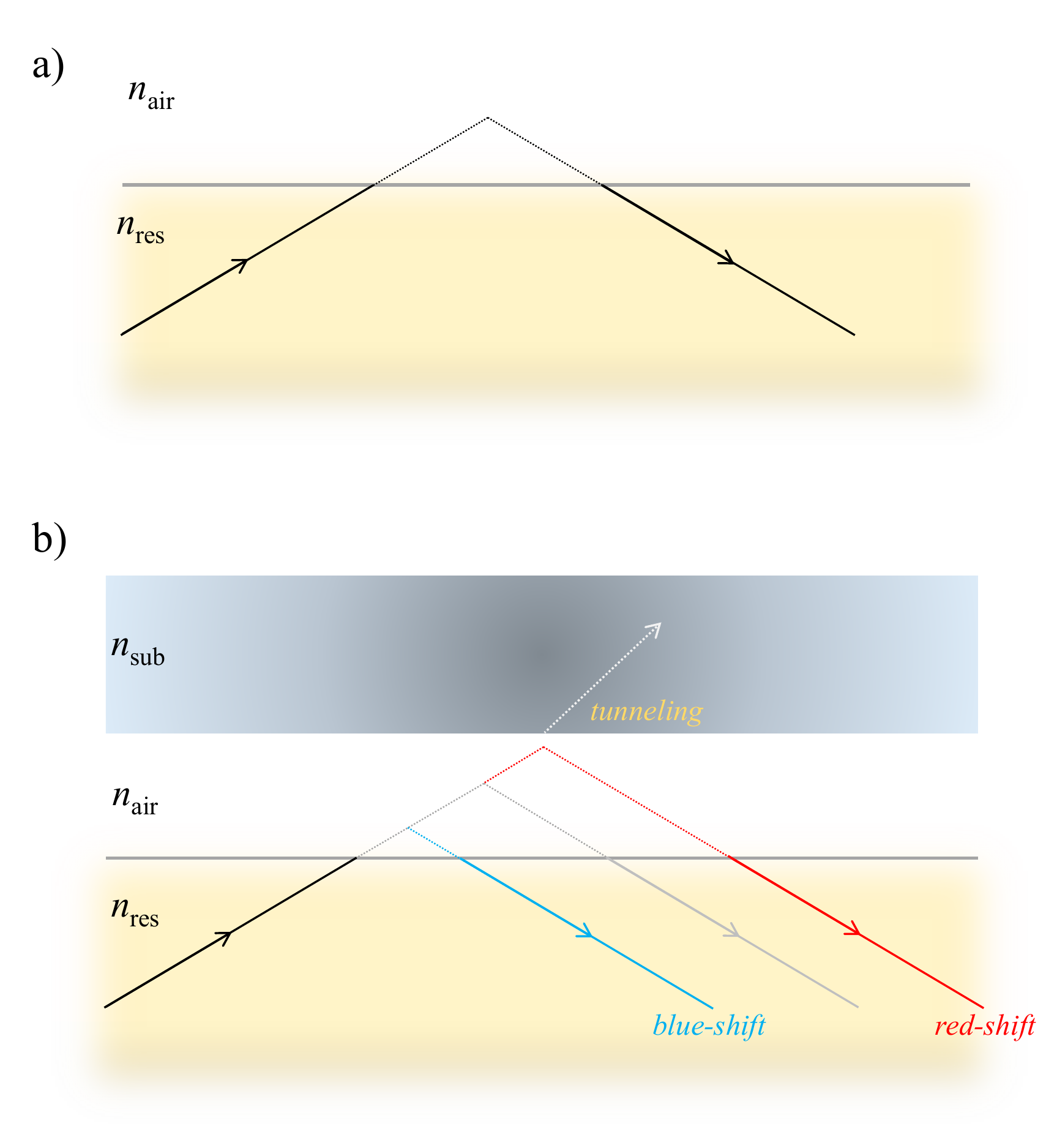}
\caption{The altered Goos-H\"{a}nchen shift at the resonator rim. a) TIR inside the resonator when a substrate is not present in its close vicinity, b) TIR inside the resonator when a substrate of refractive index $n_\text{sub}$ is introduced inside its close vicinity. This results in either a blue-shift or red-shift of light, which depends on the refractive index of the substrate.}\label{fig:Goos-Hanchen}
\end{figure}

To show that this explains the resonance shifts, we evaluate the reflection coefficient if we introduce a dielectric a small (constant) effective distance $d_\mathrm{eff}$ away from the resonator. We separate the new reflection coefficient into the terms $r = -\exp(-i(\Theta+\delta\Theta) - \alpha)$, where $\delta\Theta\in\mathbb{R}$ is the additional phase-shift due to the substrate. This phase shift should be proportional to the resonant-frequency shift, and $\alpha\geq 0$ is a loss factor due to tunneling of the light into the outer substrate, proportional to the mode broadening. $\delta\Theta$ and $\alpha$ both depend on $d_\mathrm{eff}$ and the refractive indices. The resonance shift and broadening due to the extra terms therefore are:
\begin{align}
    \text{resonance shift} &= -\frac{c \delta\Theta}{2\cos\theta_i 2\pi n R},
    \\
    \text{resonance broadening} &= 2\frac{c \alpha}{2\cos\theta_i 2\pi n R},
\end{align}
where $c$ is the speed of light in vacuum and there is a factor of two in equation (2) because the linewidth measures the power loss rate. The reflection coefficients were calculated using the {\tt tmm} Python package \cite{byrnes_multilayer_2019}.

Using the effective distance between the resonator and the dielectric as a fitting parameter, we find remarkable agreement to the theory with $d_\mathrm{eff,TE}=\SI{438.4}{\nano\meter}$ for the TE polarization for all substrate refractive indices, and $d_\mathrm{eff,TM}=\SI{352.8}{\nano\meter}$ for TM polarization, albeit only when the substrate refractive index is less than the resonator refractive index. 

A depiction of the Goos-Hänchen shift in the case of our WGM resonator and dielectric substrates is shown in Fig.~\ref{fig:Goos-Hanchen}. It shows how the resonance shift is effected by the mode being ``pushed into'' or ``pulled out of'' the resonator in the vicinity of the dielectric. The mode broadening is caused by the tunneling of photons into the dielectric.

\section{Conclusion}
In conclusion, we have observed frequency shifts and mode broadening due to the introduction of dielectric materials into the evanescent field. Remarkably, both blue- and red-shifts were observed. Our results agree qualitatively with the theory in Ref.~\cite{foreman_dielectric_2016}. The type of the shift in frequency also depends on the polarization of the excited mode. TM polarized mode is only red-shifted, whereas the TE polarized mode is both red-shifted and blue-shifted. Substrates of higher refractive index such as Ge, Si, ZnSe excite a blue-shift in the TE polarization. An anomalous blue-shift as high as \SI{14}{\mega\Hz} is observed via the Ge substrate. The red-shift is recorded at a low of \SI{-5}{\mega\Hz} when a LiNbO$_3$ substrate is used to tune the mode. These results have implications for effective mode tuning without applying any external strain which can aid in achieving the optimum phase matching conditions for non-linear processes in WGM resonators.

\section*{Funding}
N.J.L. is supported by the MBIE (New Zealand) Endeavour Fund (UOOX1805). We also acknowledge support from the Marsden Fund Grant no.~17-UOO-002. M.R.F. would like to acknowledge funding from the Royal Society.

\section*{Acknowledgments}
We would like to acknowledge fruitful discussions with Florian Sedlmeir. 

\bibliographystyle{ieeetr}
\bibliography{Dielectric}
\end{document}